# On simplified models for dynamics of pointlike objects

Marijan Ribarič and Luka Šušteršič

Jožef Stefan Institute, 1001 Ljubljana, Slovenia

E-mail:marjan.ribaric@masicom.net

## ABSTRACT

**Motivation:** We study the asymptotic-type dynamics of various real pointlike objects that one models by a variety of differential equations. Their response to an external force one defines solely by the trajectory of a single point. Its velocity eventually stops changing after cessation of the external force. The response of their acceleration to the long-term external force is slow and possibly nonlinear.

**Objective**: Our objective is to present technique for making simplified models for the long-term dynamics of pointlike objects whose motion interacts with the surroundings. In the asymptotic-type long-term dynamics, the time variable $t \in (t_m, +\infty)$ and $t_m > 0$ is large, say $t_m \nearrow +\infty$ !

**Method:** We apply Taylor series expansion to differential equations to model the acceleration of pointlike object whose response to the long-term external force is not instantaneous and possibly nonlinear.

**Results:** We make simplified models for the long-term dynamics of pointlike objects by Taylor polynomials in time derivatives of the external force.

**Application:** We interpret the relativistic Lorentz-Abraham-Dirac equation as an equation for modeling the long-term dynamics, where $t \geq t_m \gg 0$. This interpretation resolves the conceptual and usage controversy surrounding its troublesome application to determine the trajectory of a radiating charged particle, thus contributing to the development of more adequate modeling of physical phenomena.

# 1. Introduction

### A. Models in theoretical physics

The subject we consider belongs to **the modeling approach in theoretical physics**... "whose goal is to understand specific phenomena by developing either a mathematical or computational model. You begin this by choosing phenomena to study. Then you choose an approach to representing the phenomena; can you represent it as particle? a field? or some continuous distribution of matter? Then you choose a mathematical formulation. ... You then adapt your approach to the mathematical formulation, thus developing a mathematical representation of your phenomena. You then use physical, mathematical, and/or computational arguments and methods to make predictions in the form of tables, plots, and/or formulas. By studying these results in different circumstances, you can extend our understanding of the phenomena. This is the most direct method of doing theoretical physics; it is a straight application of mathematical or computational methods. It is certainly the most structured way of doing theoretical physics. ... A body of models linked by physical argument, derivation methods, and/or computer simulations constitute a physical theory.... The first step in understanding any physics is to try to simplify the situation by removing all complications and then by working out all of the consequences of the situation. The particle is this kind of simplification. For such a simple explanation, it is very rich in principles and consequences. Once we have studied many simple ideas, we need to make them more realistic by reintroducing some of the complications that we removed in the process of simplification." see G. E. Hrabovsky [1].

**Feynman's advice** [2]:"I think equation guessing might be the best method to proceed to obtain the laws for the part of physics which is presently unknown... the problem is not to find the best or most efficient method to proceed to a discovery, but to find any method at all."

Thus, we put forward the method for making **simplified**, **serial models** (**S-models**):
**i)** Considering a phenomenon, we chose the fundamental model that we will simplify;
**ii)** modifying it by inserting an **auxiliary parameter** for expansion by the Taylor series,
**iii)** the S-models we make up of the Taylor polynomials, omitting auxiliary parameter.
There are three illustrative examples of such simplified, serial modeling:

**a)** Given physical phenomenon described by a complicated scalar function $f(t)$, we insert two auxiliary parameters $\lambda$ and $t_0$ , and use the Taylor series

$$f(\, t_0 + \lambda(t - t_0)) = \sum_{n=0}^{\infty} f^{(n)}(t_0\,)[\lambda(t - t_0)]^n/n! \qquad \text{(I.1)}$$

with $\lambda = 1$ to formulate the following two-term S-model

$$f_l(t) \equiv f(t_0) + f^{(1)}(t_0)(t - t_0). \qquad \text{(I.2)}$$

This linear model is simpler than $f(t)$ and suggests two, yet simpler one-term models:

$$f_c(t) \equiv f(0) \quad \text{and} \quad f_p(t) \equiv f^{(1)}(0)\, t\,. \qquad \text{(I.3)}$$

- Note that $f(t)$ may not be equal to its Taylor series even if the series converges!

**b**) Given the convolution integral

$$y(t) = \int_{-\infty}^{\infty} f(t')x(t-t')\,dt', \qquad \text{(I.4)}$$

we use the Taylor series expansion of $x(t-\lambda t')$ . The first m+1 terms provide the S-model

$$y_m(t) \equiv \sum_0^m a_n\, x^{(n)}(\mathrm{t}), \qquad \text{(I.5)}$$

where $a_n$ are constants that are defined by the moments of $f(t)$ specified by equation (I.4).

**c)** We can interpret the partial differential equations as the simplified serial models obtained from the Boltzmann integro-differential transport equations, see [3, Sect.4.3.1].

❖ As noted by Dirac, "A great deal of my work is just playing with equations and seeing what they give."..." A good deal of my research in physics has consisted in not setting out to solve some particular problem, but simply examining mathematical equations of a kind that physicists use and trying to fit them together in an interesting way, regardless of any application that the work may have. It is simply a search for pretty mathematics. It may turn out later to have an application. Then one has good luck." see [5].

❖ Simplified, serial models are convenient for detailed study of phenomena. Selecting a particular element of the fundamental model for inserting the auxiliary parameter, we concentrate our attention on the specific aspect of the considered phenomenon.

❖ A wide variety of fundamental models implies the same S-model! Thus, one can collate a variety of simplified models without knowing the actual fundamental model.
It makes sense to consider initially various simplified models for a phenomenon in order to infer eventually its fundamental model, see [4].

## B. Subjects

In 2012, we presented modeling of the long-term, asymptotic dynamics of a point-like object where $t \in (t_m, +\infty)$ and $t_m$ is large, (arXiv:1205.2920 [physics.gen-ph]). There we introduced some elements and concepts that we will now clarify with simple examples. In this paper, we consider simplified, serial models of the long-term dynamics, of a particular type of rigid bodies whose motion interacts with the surrounding medium. We name them "pointlike objects" (POs) and define them:

**a)** A **PO** is a classical extended real object that is moving through a medium. The response of PO to an external force is specified solely by the trajectory of a single point, which we name "the PO-position". Therefore, the POs are a sort of rigid bodies.

**b)** The external force determines the acceleration of the PO-position.

**c)** The PO-velocity eventually stops changing after the cessation of the external force.

**d)** So far, one has modeled the dynamics of a PO by using so-called "point mass", whose acceleration one has specified by Newton's second law of motion, dividing the force acting on the point mass by its mass, its sole kinetic constant. However, a PO may be actually a very complicated real object such as a satellite, train, ship, chain, particle . . . , thus we may oversimplify the situation by using a point mass to model its dynamics. Therefore, we approach this question by considering simplified serial models of the long-term, PO dynamics (**LT-dynamics** for short) for a small and slowly changing external force $\lambda F(\lambda \mathrm{t})$ with a variable auxiliary parameter $\lambda \geq 0$. Expanding the LT-dynamics in powers of λ, we obtain with $\lambda = 1$ simplified, serial models: the Taylor polynomials in terms of the external force $\lambda F(\lambda \mathrm{t})$ and its derivatives, which we name "**the LT-formulas**".

In **Section 2**, we consider the damped harmonic oscillator driven by the small and slowly changing external force $\lambda F(\lambda t)$ with auxiliary parameter $\lambda$ in order to calculate illustrative S-models for LT-dynamics where $t \in (t_m, +\infty)$ and $t_m$ is large, say $t_m \nearrow +\infty$ .As the fundamental model, we introduce a special type of Newton's second law formula for acceleration of PO, which we name **the N-formula,** where $t \in (0, +\infty)$. It generalizes Newton's second law by explicitly specifying the PO-acceleration by a possibly nonlinear transform of the external force. We calculate the N-formula for the driven damped harmonic oscillator from its differential equation of motion. By the expansion of this N-formula in powers of $\lambda$, we obtain the corresponding LT-formulas, the Taylor polynomials in time derivatives the external force $\lambda F(\lambda t)$. These LT-formulas suggest that by eliminating iteratively the higher time derivatives of the trajectory from the PO differential equation of motion, we can calculate the corresponding LT-formulas up to any order of $\lambda$, without solving the differential equation.

In **Section 3**, we consider S-models for the LT-dynamics of PO whose motion interacts with the surrounding medium. We calculate iteratively exemplary LT-formulas without solving the relevant differential equations: i) the Riccati differential equation for velocity at the quadratic drag force, ii) a second order cubic nonlinear differential equation for the driven damped oscillator, and iii) the Lorentz-Abraham-Dirac relativistic differential equation.

In **Section 4**, we draw coclusion and comment on:

a) Specific S-models for the LT-dynamics.

b) Dynamic properties of PO, which serial models for relativistic LT-dynamics imply.

c) Applications of LT-formulas for predicting the long-term acceleration dependence of PO on the external force and its time derivatives: this dependence determines the relative significance of kinetic constants of the PO equation of motion for the LT-dynamics.

d) Figure 1 shows the block cycle diagram that illustrates the relations between the introduced concepts and techniques.

❖ Since physical theories employ mathematical models to describe and predict physical phenomena, our knowledge depends on the models available to that end. To increase their scope we use the Taylor series expansion in order to introduce simplified, serial models. Using the first few terms of a Taylor series, these models provide some information about difficult and complex problems. In this paper, we use them to describe the long-term dynamics of a special type of rigid bodies at small and slowly changing external forces.

## C. List of Contents

## 2. LT-dynamics of the driven damped harmonic oscillator

To illustrate the mathematical framework of the proposed S-models for of the LT-dynamics, we will now consider the S-models for the LT-dynamics of the damped harmonic oscillator that is driven by an external force $\lambda F(\lambda t)$ with its magnitude and rate of change determined by the auxiliary parameter $\lambda > 0$.

### 2.1 N-formula for the driven damped harmonic oscillator

We calculate the N-formula from the differential equation of motion for the particular kind of PO. This PO is based on the point mass that is moving along the x-axis under the influence of the external force $f(t)$, $f(0) = 0$, and initially at rest at $x(0) = 0$ . The point mass is attached to the zero-length spring with the force constant $k \geq 0$ , and slowed down by the frictional force with the non-negative viscous damping coefficient $c$ such that $c > 0$ when $km \neq 0$. Thus, the PO-position $x(\mathrm{t})$ satisfies $\forall$ t $\geq 0$ the differential equation of motion for the driven damped harmonic oscillator:

$$m\, x^{(2)} + \mathrm{c}\, x^{(1)} + k\, x \;= f \quad \text{with} \quad x^{(n)} \equiv (d/dt)^n\, \mathrm{x}\,, \;\; \mathrm{n} \;=\; 0, 1, 2 \ldots\,, \qquad (1)$$

where the kinetic constant $m$ specifies the PO inertial force, whereas the kinetic constants $c$ and $k$ specify the interaction between the surrounding medium, and PO-velocities' and PO-position respectively. Therefore the PO-trajectory

$$x(t) = \int_0^t z\,(\mathrm{t}')f(\mathrm{t}-\mathrm{t}')\, d\mathrm{t}' \quad \text{if} \;\; m,\, c,\, \text{and}\; k > 0 \qquad (2)$$

with the Green function

$$z(t) \;\equiv \left(m\sqrt{1-\zeta^2}\,\omega_0\right)^{-1} \exp(-\zeta\omega_0 \mathrm{t})\, \sin(\sqrt{1-\zeta^2}\,\omega_0 \mathrm{t}) : \qquad (3)$$

$\omega_0 = \sqrt{k/m}$ is named "the un-damped angular frequency", and $\zeta = c/2m\omega_0$ is named "the damping ratio". The PO-acceleration as a function of the external force $f(t)$ is given by the following linear N-formulas:

$$x^{(2)}(\mathrm{t}) = \mathrm{m}^{-1} f(\mathrm{t}) - \mathrm{m}^{-1} \int_0^{\mathrm{t}} z\,(\mathrm{t}')[\,kf(\mathrm{t}-\mathrm{t}') \;+ cf^{(1)}(\mathrm{t}-\mathrm{t}')]d\mathrm{t}' \quad \text{if} \;\; m,\, c,\, \text{and}\; k > 0\,; \quad (4)$$

$$x^{(2)}(\mathrm{t}) = \; m^{-1}\, f(\mathrm{t}) - \; c\, m^{-2} \int_0^{\mathrm{t}} \exp(-c\mathrm{t}'/m)\, f(\mathrm{t}-\mathrm{t}')\, d\mathrm{t}' \quad \text{if}\; m > 0,\; c \geq 0,\; k = 0\,; \qquad (5)$$

$$x^{(2)}(\mathrm{t}) = c^{-1} f^{(1)}\,(\mathrm{t}) - k\, c^{-2} \int_0^{\mathrm{t}} \exp(-k\mathrm{t}'/c)\, f^{(1)}(\mathrm{t}-\mathrm{t}')\, d\mathrm{t}' \quad \text{if}\; m = 0,\; c > 0,\; k \geq 0\,; \qquad (6)$$

They transform the external force into PO-acceleration by integral operators; cf. the N-formulas (16), (17), and (18).

When the external force $f(t) = 0 \;\; \forall \;\; t \geq t_1$, the PO equation of motion (1) implies:

$$x(\mathrm{t}) = a_1 \exp(-\zeta\omega_0 \mathrm{t}) \sin(\sqrt{1-\zeta^2}\omega_0 \mathrm{t} + \varphi) \quad \text{if} \;\; \mathrm{m, c\,, k} \;> 0\,; \qquad (7)$$

$$x^{(1)}(\mathrm{t}) = a_2 \exp(-ct/m) \quad \text{if} \;\; m > 0,\; c \geq 0,\; k = 0\,; \qquad (8)$$

$$x(\mathrm{t}) = a_3 \exp(-k\mathrm{t}/c) \quad \text{if} \;\; m = 0,\; c > 0,\; k > 0\,; \qquad (9)$$

where the four constants $a_1$, $a_2$, $a_3$, and $\varphi$ are determined by $f(t)$, $0 < \mathrm{t} < \mathrm{t}_1$.

Thus, after the cessation of the external force $f$(t) the PO-velocity $x^{(1)}(t)$ eventually stops changing, and the properties of the external force $f$(t) within every finite period of time have negligible effects on the PO-velocity $x^{(1)}$(t) as $t \to +\infty$ because the PO differential equation of motion (1) is linear. If c < 0, then there is self-acceleration.

❖ To further illustrate *the relationship* between linear *N-formulas* and linear PO equations of motions we calculate *N-formula* for a PO made out of **two connected point masses** of equal mass m $\geq 0$, which are located on the x-axis, initially resting at points $x(0) = 0$ and $x_1(0) = 0$: thus $x^{(1)}(0) = 0$ and $x_1^{(1)}(0) = 0$. They are *connected* by the zero-length spring with the force constant k/2 > 0. The point mass with the trajectory x (t) is accelerated by the external force $f$(t), $f(0) = 0$, and slowed down by the frictional force $-c\,x^{(1)}$(t) with the viscous damping coefficient $c \geq 0$. Whereas the point mass with the trajectory $x_1$(t) is only slowed down by the frictional force $-c\,x_1^{(1)}$(t) . The differential equations of motion for this system of two connected point masses are

$$mx^{(2)} = -cx^{(1)} + f - ½\,k\,(x - x_1) \quad \text{and} \quad mx_1^{(2)} = ½\,k\,(x - x_1) - cx_1^{(1)}. \quad (10)$$

Therefore, the velocity of the first point mass is given by the formula:

$$x^{(1)}(\mathrm{t}) = \int_0^{\mathrm{t}} \mathrm{z}(\mathrm{t}-\mathrm{t}')\,[f^{(1)}(\mathrm{t}') + \mathrm{k}/(2\mathrm{m}) \int_0^{\mathrm{t}'} \exp(-c\tau/m)\, f(\mathrm{t}'-\tau)d\tau]d\mathrm{t}' \ \text{ if } m > 0, c \geq 0; \ (11)$$

and

$$x^{(1)}(\mathrm{t}) = c^{-1} \int_0^{\mathrm{t}} \exp(-k\mathrm{t}'/c)\,[\mathrm{k}/(2\mathrm{c})\, f(\mathrm{t}-\mathrm{t}') + f^{(1)}(\mathrm{t}-\mathrm{t}')\,]d\mathrm{t}' \ \text{ if } m = 0, c > 0. \quad (12)$$

If $c > 0$, we may consider $x$(t) as the PO-trajectory, and the time differentiation of the formula (11) or (12) defines directly the corresponding N-formula. According to the differential equations of motion (10), the differential equation of motion for the PO-trajectory $x$(t) is as follows:

$$m^2x^{(4)} + 2cm\,x^{(3)} + (km + c^2)\,x^{(2)} + kc\,x^{(1)} = ½k\,f + c\,f^{(1)} + m\,f^{(2)}\ ,\ c > 0\,. \quad (13)$$

The kinetic constants $m$ and $k$ specify the PO, whereas the kinetic constant $c$ specifies the interaction between the PO-velocity and the surrounding medium.

## 2.2 S-models for the linear LT-dynamics

We now calculate approximate models for the linear LT-dynamics implied by the linear N-formulas (4) and (5), when the external force $f(\mathrm{t}) = \lambda F(\lambda \mathrm{t})$ i.e. $\boldsymbol{\lambda} > 0$ is small and $t \in (t_m, +\infty)$ . Taylor series expansion of N-formulas (4) and (5) in powers of λ gives:

$$x^{(2)}(\mathrm{t}) = k^{-1}\lambda\, F^{(2)}(\lambda \mathrm{t}) - k^{-2}c\,\lambda\, F^{(3)}(\lambda \mathrm{t}) + \mathrm{O}(\lambda^5) \quad \text{as } t_m \to +\infty \quad \text{if } m,c,k > 0\,; \quad (14)$$

and

$$x^{(2)}(\mathrm{t}) = c^{-1}\lambda\, F^{(1)}(\lambda \mathrm{t}) - mc^{-2}\lambda\, F^{(2)}(\lambda \mathrm{t}) + \mathrm{O}(\lambda^4) \quad \text{as } t_m \to +\infty \quad \text{if } m,c > 0,\ k = 0\,; \quad (15)$$

provided $sup_{t \geq 0}\left| F^{(n)}(\lambda \mathrm{t})\right| \leq \infty$ for n = 0, 1, 2, 3. Thus, both responses of acceleration to the external force are slow, not instantaneous.

Whereas the differential equation of motion (1) implies that $\forall$ t $\geq 0$ and each $\lambda$ :

$$x^{(2)}(\mathrm{t}) = m^{-1}\lambda\, F(\lambda t) \quad \text{if } m > 0,\ c = 0,\ k = 0\,; \quad (16)$$

$$x^{(2)}(\mathrm{t}) = c^{-1}\lambda F^{(1)}(\lambda t) \quad \text{if } m = 0,\ c > 0,\ k = 0\,; \quad (17)$$

$$x^{(2)}(\mathrm{t}) = k^{-1}\,\lambda F^{(2)}(\lambda t) \quad \text{if } m = 0,\ c = 0,\ k > 0\,. \quad (18)$$

Above formulas (14) and (15) show how using the S-models for the LT-dynamics specified by the PO equation of motion (1) with $f(t) = \lambda F(\lambda t)$, we can obtain the long-term PO-acceleration expressed as a sum of time derivatives of the small and slowly changing external force $\lambda F(\lambda t)$ , say,

$$x^{(2)}(t) = \sum_1^N k_n \lambda F^{(n-1)}(\lambda t) + O(\lambda^{N+1}) \quad (19)$$

provided $sup_{t\geq 0}\left| F^{(n)}(\lambda t)\right| \leq \infty$ for n = 0, 1,...,N; e.g. if $k_1 = k^{-1}$, $k_2 = k^{-2}c$ $with\, m, c, k > 0$. Such S-model for the LT-dynamics under small and slowly changing external force $\lambda F(\lambda t)$ we name as **the LT-formula of the order of $\lambda^N$.** The real constants $k_n$ we name as **the LT-constants**. About the LT-formula (19) we note that

$$x^{(n)}(t) = O(\lambda^{n-1}) \text{ as } t \to +\infty \quad , n = 2, 3, 4, \ldots . \quad (20)$$

If $k_n = 0$ for $n = 1, 2, \ldots, o-1$ , and $k_o \neq 0$, o ≥ 1, then each LT-formula (19) of the order $\lambda^N$ implies iteratively the following novel differential equation of the order $\lambda^{o+N-2}$ for the LT-dynamics: **the LT-equation**

$$\sum_2^N c_n x^{(n)}(t) = \lambda F^{(o-1)}(\lambda t) + O(\lambda^{o+N-1}) , \quad (21)$$

and vice versa. When N = 4 and $k_1 \neq 0$, the constants of the LT-equation (21) and the LT-constants of the LT-formula (19) are related iteratively:

$$c_2 = k_1^{-1}, \; c_3 = -k_1^{-2}k_2 , \text{ and } c_4 = k_1^{-3}k_2^2 - k_1^{-2}k_3 ; \quad (22)$$

$$k_1 = c_2^{-1}, \; k_2 = -c_3 c_2^{-2} , \text{ and } k_3 = c_2^{-3}c_3^2 - c_2^{-2}c_4 . \quad (23)$$

## 2.3 On elements and usage of serial models

**1)** The equations (1), (4), (14), (19), and (21) illustrate the mathematical relations between the following elements of the proposed S-models :

**a)** PO differential equation of motion, **b)** N-formula, **c)** LT-dynamics, **d)** LT-formulas, and **e)** LT-equations; see Figure 1 the block cycle diagram in Section 4.

**2)** The LT-formulas (19) contain *only one* time derivative of the PO-trajectory, the LT-equations (21) contain *only one* time derivative of the external force, while the PO differential equations of motion are is not so constrained, cf. equation (13).

**3)** The equations (14) and (15) exemplify how on expresing the LT-constants $k_n$ of the LT-formula (19) in terms of the kinetic constants *m*, *c*, and *k,* the S-models bring out the relative significance of these kinetic constants for the LT-dynamics.

**4)** The early time, $t \downarrow 0$, start-up dynamics of the N-formula (4) is given by

$$x^{(2)}(t) = m^{-1}\lambda^2 F^{(1)}(0)\, t + O(\lambda^3 t^2) . \quad (24)$$

But the way in which the LT-formula (14) of the LT-dynamics depends on the kinetic constants *m*, *c*, and *k* is fundamentally different.

**5)** The equations (4) and (14) suggest that the N-formula and the corresponding LT-formulas may differ significantly. Nevertheless, when the external force is small and slowly changing, we may use the LT-formula to calculate the approximate long-term PO-trajectories.

**6)** In contrast to the differential equation of motion (1) that depends continuously on the kinetic constants m*, c,* and *k*, in view of the equations (14) and (15), the corresponding LT-formulas may not. The same applies for the LT-equations (21),
which in general do not determine the PO differential equation of motion (1).
**7)** The equations (15) to (18) and the estimate (20) suggest that without solving a PO equation of motion like (1) or (13), we can calculate the corresponding LT-formulas up to any order of $\lambda$ by eliminating iteratively the higher time derivatives of trajectory. So assuming that the estimate (20) is correct, we calculate in that way from the differetial equation of motion (13) for the PO consisting of two connected point masses the correspondig LT-formula:

$$x^{(1)}(\mathrm{t}) = (2\mathrm{c})^{-1}\lambda\, F(\lambda \mathrm{t}) + O(\lambda^2) \quad \text{if } \ m \ \geq \ 0, \ c \ > 0 \qquad (25)$$

**8)** The formulas (7) to (9) show that for the particular PO, its initial response to an external force becomes exponentially negligible. Moreover, if the PO-acceleration is cyclic then there is no start-up, early time dynamics: in this case, the long-term dynamics and dynamics of PO are identical!

## 3. Exemplary LT-formulas

In this section, where $t \in (t_m\,, +\infty)$ with $t_m \nearrow +\infty$, we calculate iteratively from the assorted differential equations some exemplary LT-formulas in terms of the time derivatives of a function of the external force $\lambda F(\lambda \mathrm{t})$. Any LT-formula or LT-equation that we do not calculate from some PO equation of motion we designate as a *hypothetical* one.

### 3.1 LT-formulas for the strong string

Using the strong string, we generalize the PO differential equation of motion (1):

$$m\, x^{(2)} + \mathrm{c}\, x^{(1)} + \ k_1\, x + k_3 \mathrm{x}^3 \ = \ \lambda F(\lambda \mathrm{t}) \qquad (26)$$

where $m,$ c, $k_1,\ k_3 \geq 0;$ and $\mathrm{c} > 0 \ \ if \ \ (k_1 + \ k_3)m > 0.$ Thus the three kinetic constants c , and $k_1$ and $k_3$ , specify the reaction of the surrounding medium to the PO-velocity and PO-position respectively. To calculate the Taylor polynomial of the differential equation (26) we rewrite it by Cardano's formula:

$$\boldsymbol{x} \ = \left[\, -q + (\, q^2 + p^3)^{1/2} \ \right]^{1/3} - \ \left[q + (\, q^2 + p^3)^{1/2} \ \right]^{1/3} \qquad (27)$$

with $\ p \ \equiv k_1/3k_3\,,\ q \ \equiv \ \left(\mathrm{r} - \lambda F(\lambda \mathrm{t})\right)/2k_3\,,$ and $r \equiv m\, x^{(2)} + c\, x^{(1)}$ . We presume that the equation (26) is still a PO equation of motion if $\ k_3 > 0$ and that the estimates

$$x^{(\mathrm{n})}(\mathrm{t}) = \mathrm{O}\left(\lambda^{\mathrm{n}+2\mathrm{sig}(k_1)/3+1/3}\right) \ \text{as} \ \ t \ \to +\infty \ , \ \ \mathrm{n} = 0, 1, \ldots, \qquad (28)$$

are true also if $\ m, c > 0$, calculate the Taylor polynomial of $\boldsymbol{x} \equiv$ [rhs(27)] as a function of *r;* and eliminate the time derivatives $x^{(\mathrm{n})}$ . This way we obtain the LT-formula:

$$x = \boldsymbol{x}^{\{\mathbf{0}\}} + c\, \boldsymbol{x}^{\{\mathbf{1}\}}\left(\boldsymbol{x}^{\{\mathbf{0}\}}\right)^{(1)} + \mathrm{O}\left(\lambda^{7/3+2\mathrm{sig}(k_1)/3}\right) \ \text{as} \ \ t \ \to +\infty\,,$$

$$\text{where} \ \ \boldsymbol{x}^{\{\mathbf{n}\}} \equiv (\partial/\partial r)^{\mathrm{n}}[\mathrm{rhs}(27)] \text{ at } r = 0\,. \qquad (29)$$

This formula exemplifies a new type of S-models for the LT-dynamics with polynomials in time derivatives of the function $\boldsymbol{x} \equiv$ [rhs(27)] of the external force $\lambda F(\lambda \mathrm{t})$.

### 3.2 The quadratic drag force

Let us consider a PO with mass m ≥ 0, which is moving along the x-axis through a fluid at relatively large velocity $x^{(1)}(t) > 0$ under the influence of the external force $\lambda F(\lambda t) > 0$, and slowed down by the Lord Rayleigh type of the quadratic drag force $c_d(x^{(1)}(t))^2$, $c_d > 0$. The PO-mass $m$ specifies the PO inertial force, whereas the constant $c_d$ specifies the reaction force of the surrounding fluid. So we presume that the PO-velocity $x^{(1)}$ satisfies the following Riccati differential equation:

$$m\, x^{(2)} + c_d (x^{(1)})^2 = \lambda F(\lambda t)\,. \qquad (30)$$

Hypothesizing that the differential equation (30) is a LT-equation and that the estimates

$$x^{(n)}(t) = O\big(\lambda^{n-1/2}\big) \quad \text{as } t \to +\infty\,, \ n = 1, 2, \dots\,, \qquad (31)$$

are true also for $m > 0$, we get iteratively a new type of serial models for LT-dynamics. They are polynomials in time derivatives of $\sqrt{\lambda F(\lambda t)}$:

$$x^{(1)}(t) = \sqrt{\lambda/c_d}\,\sqrt{F(\lambda t)} - m/4\,\sqrt{\lambda/c_d}\,F^{(1)}(\lambda t)/\sqrt{F(\lambda t)} \; + \; O\big(\lambda^{5/2}\big) \quad \text{as } t \to +\infty\,. \quad (32)$$

We can generalize the calculated hypothetical LT-formula (32) by adding the frictional force $-c\, x^{(1)}(t)$ to the hypothetical LT-equation (30).

### 3.3 The nonlinear relativistic LT-formulas

Let us consider LT-formulas of a relativistic N-formula. We base them on relativistic point mass, which is located at $\boldsymbol{r}(t)$ and moving with velocity $\boldsymbol{v}(t)$, under the influence of the external force $\lambda \boldsymbol{F}(\lambda t)$ with the dimensionless auxiliary parameter $\lambda > 0$. We define the external four-force

$$\lambda \Phi(\lambda t) \equiv \gamma\big(\boldsymbol{\beta}(t)\cdot \lambda \boldsymbol{F}(\lambda t),\ \lambda \boldsymbol{F}(\lambda t)\big) \quad \text{with} \quad \gamma(t) \equiv (1 - |\boldsymbol{\beta}|^2)^{-1/2}\,, \qquad (33)$$

where $\boldsymbol{\beta}(t) \equiv \boldsymbol{v}/c$. We will use the PO four-velocity $\beta(t) \equiv (\gamma, \gamma\boldsymbol{\beta})$; and the metric with the signature $(+ - - \ -)$, so that $\beta \cdot \beta = 1$. We introduce an additional four-force $\Delta(t)$, which specifies the properties of PO-dynamics in the case of the external four-force $\lambda\Phi(\lambda t)$, and formulate the relativistic N-formula:

$$mc\beta^{[1]}(t) = \Delta(t) + \lambda\Phi(\lambda t) \quad \text{with} \quad \beta^{[n]} \equiv (\gamma d/dt)^n \beta\,, \quad n = 0, 1, 2, \dots\,, \qquad (34)$$

where $t/\gamma$ is the proper time. As $\beta \cdot \beta^{[1]} = 0$ and $\beta \cdot \Phi = 0$, we may rewrite the relativistic N-formula (34) as

$$mc\beta^{[1]}(t) = (1 - \beta\,\beta\,\cdot)\Delta(t) + \lambda\Phi(\lambda t)\,. \qquad (35)$$

Generalizing the linear LT-formula (19), we model the dependence of the four-force $\Delta(t)$ on the external four-force $\lambda\Phi(\lambda t)$ by a relativistic polynomial in time derivatives $\lambda\Phi^{[n]}$ of the external four-force, to get the hypothetical relativistic LT-formula:

$$\begin{aligned}\beta^{[1]} = (1 - \beta\,\beta\,\cdot)\big[&k_1\,\lambda\Phi + k_2\,\lambda\Phi^{[1]} + k_{31}\lambda^3(\Phi\cdot\Phi)\,\Phi + k_{32}\,\lambda\Phi^{[2]} + k_{41}\,\lambda^3(\Phi^{[1]}\cdot\Phi)\Phi \\ &+ k_{42}\,\lambda^3\big(\Phi\cdot\Phi\big)\Phi^{[1]} + k_{43}\,\lambda\Phi^{[3]} + k_{51}\lambda^5\,(\Phi\cdot\Phi)^2\,\Phi + k_{52}\,\lambda^3\big(\Phi^{[1]}\cdot\Phi^{[1]}\big)\Phi \\ &+ k_{53}\,\lambda^3(\Phi^{[1]}\cdot\Phi)\Phi^{[1]} + k_{54}\,\lambda^3(\Phi^{[2]}\cdot\Phi)\Phi + k_{55}\,\lambda^3(\Phi\cdot\Phi)\Phi^{[2]} + k_{56}\,\lambda\Phi^{[4]}\big] + O(\lambda^6)\,,\end{aligned} \qquad (36)$$

where the real parameters $k_1, \dots, k_{56}$ are independent of the external four-force $\lambda\Phi(\lambda t)$. We name them "the LT-constants" when we use them for a particular PO to specify its hypothetical relativistic LT-formula (36) up to the order of $\lambda^5$ inclusive.

• If $\Delta(\mathrm{t}) = \Delta(-\mathrm{t})$, then N-formula (35) is invariant under time reversal and the hypothetical relativistic LT-formula (36) has $k_2 = k_{41} = k_{42} = k_{43} = 0$.
• The N-formulas (17) and (18) themselves suggest two hypothetical relativistic LT-formulas: $[\mathrm{rhs}(36)]^{[1]}$ and $[\mathrm{rhs}(36)]^{[2]}$.
• By eliminating iteratively all the time derivatives $\lambda\Phi^{[n]}$ except the one of $\lambda\Phi^{[0]}$ from the hypothetical relativistic LT-formula (36), we get a hypothetical relativistic LT-equation, analogous to the LT-equation (21).

### 3.4 The relativistic LT-formulas for an electrified PO

Presuming that the PO is electrified by a pointlike charge, we follow Schott [6] and express the additional four-force $\Delta(\mathrm{t})$ as the difference

$$\Delta = -\,d\big(\beta^{[1]}\cdot\beta^{[1]}\big)\,\beta \;+\; B^{[1]}\,, \tag{37}$$

where $d\big(\beta^{[1]}\cdot\beta^{[1]}\big)\,\beta$, $d \geq 0$, is the long-term intensity of the four-momentum emitted by the Liénard-Wiechert potentials with the *cyclically* moving singularity at $\boldsymbol{r}(t)$, see [7, §6.6], and $B^{[1]}$ is the time derivative of an "acceleration four-momentum $B(t)$". Therefore, for an electrified PO we rewrite the relativistic N-formula (34) as

$$mc\beta^{[1]} - d\big(\beta^{[1]}\cdot\beta^{[1]}\big)\,\beta \;+\; B^{[1]} \;= \lambda\Phi\,, \tag{38}$$

and

$$\beta\cdot\big(B + d\beta^{[1]}\big)^{[1]} \;=\; 0\;. \tag{39}$$

The kinetic constant $d$ specifies the magnitude of radiation reaction force of the surrounding vacuum, which opposes the acceleration of an electrified PO. According to Dirac [8], the acceleration four-momentum $B(t)$ may be any four-function of the time derivatives $\beta^{[\mathrm{n}]}$; and the N-formula (38) for an electrified PO conserves the four-momentum by the equation (39). Furthermore, Bhabha [9] pointed out that if $B(t)$ is such that the cross product

$$\beta \wedge \big(B + d\beta^{[1]}\big) \tag{40}$$

is a total differential with respect to the proper time, then such an electrified PO conserves the angular four-momentum !

Inspired by the LT-equation (21), we assume that the nth time derivative $\beta^{[\mathrm{n}]}$ is of the order $\lambda^n$ as $t \to +\infty$, and then model the time derivative $B^{[1]}$ in the relativistic N-formula (38) by the relativistic polynomials in $\beta^{[\mathrm{n}]}$, subject to the Bhabha condition (40), cf. [7, Ch.9] and [10]. Accordingly, the hypothetical, relativistic LT-equation for an electrified PO equals the LT-equation

$$mc\beta^{[1]} - d(1 - \beta\,\beta\,\cdot)\beta^{[2]} = \lambda\Phi \tag{41}$$

up to the order of $\lambda^2$ inclusive, disregarding the Bhabha condition (40). Assuming the Bhabha condition (40), we gave such a hypothetical, relativistic LT-equation up to the order of $\lambda^6$ inclusive, see [11]. On eliminating iteratively $\beta^{[2]}$ from the (41) we get

$$mc\beta^{[1]} = (1 - \beta\,\beta\,\cdot)\big[\lambda\Phi + \lambda d/mc\,\Phi^{[1]}\,\big] + O(\lambda^3)\,, \tag{42}$$

which is the LT-equation such as the LT-equation (36) up to the order of $\lambda^2$. Up to the order of $\lambda^6$inclusive, we gave such relativistic LT-equation in [7, Sect.11.4].

### 3.5 On equation of motion for a charged particle

**1)** In 1892, H. A. Lorentz started an ongoing quest to take account of the radiation reaction force (the effect of the loss of four-momentum by the electromagnetic radiation) by the classical equation of motion for a charged particle. In 1938, Dirac assumed that an electron is such a simple thing that the lowest order hypothetical LT-equation (41) ought to be the correct equation of motion with $d = e^2/6\pi\epsilon_0 c^2$ , and no additional polynomial terms are needed, cf. [8]. The equation (41) is named the Lorentz-Abraham-Dirac equation. Since it exhibits the self-acceleration, it has baffled the mathematical physicists ever since the Dirac invented it. Since we obtained it as the first known hypothetical relativistic LT-equation about the LT-dynamics of an electrified PO, we puzzled out its significance. *An equation about long-term dynamics,* where $t \in (t_m, +\infty)$, *is not necessarily an equation of motion* ,where $t \in (0, +\infty)$*, the initial dynamics is missing.*

**2)** In 2008, Rohrlich [12] stated that the physically correct equation of motion for a classical charged particle is the lowest order relativistic LT-formula (42) for an electrified PO with $d = e^2/6\pi\epsilon_0 c^2$, provided $\left|(d/mc)(1 - \beta\,\beta\,\cdot)\Phi^{(1)}\right| \ll |\,\Phi|$ , see [13] for comment.

There is a century old discussion with an infinite number of proposals about the appropriate equation of motion for an electrified PO, see e.g. [7, 14, 15] and the references cited therein. Nevertheless, we do not consider here the theoretical problem about the properties of a real object and conditions under which it behaves largely like PO, and we may idealize it as PO by constructing an appropriate PO equation of motion.

## 4. Conclusion

### A. Comments

### 4.1 Multi-simplified serial model.

To simplify an implicit relation $\mathcal{R}(x_1, x_2, c) = 0$ we modify it by inserting six parameters:

$$\mathcal{R}(\lambda_1 x_1 + \varepsilon_1,\ \lambda_2 x_2 + \varepsilon_2,\ \lambda_c c + \varepsilon_c) = 0, \qquad (43)$$

where three auxiliary parameters $\lambda_1, \lambda_2, and\ \lambda_c$ point to the variables $x_1$ , $x_2$, and the constant $c$ respectively, whereas the parameters $\varepsilon_1$, $\varepsilon_2$, and $\varepsilon_c$ specify their starting values. The Taylor series to nth order in auxiliary parameters $\lambda_1, \lambda_2, and\ \lambda_c$ provides the nth order multi-simplified, serial model.

### 4.2 Specific S-models for LT-dynamics

Usage of S-models requires selecting and identifying the relevant aspects of the LT-dynamics. Instead of using the modified external force $\lambda F(\lambda t)$, we might obtain more detailed information about the significance of kinetic constants for LT-dynamics of a given PO by using the modified external force $\lambda F(\varepsilon t)$ with two dimensionless, auxiliary parameters λ and ε, to control separately the magnitude and the rate of change of the external force. Considering the same PO, we can start also with various simplified equations of motion. Thereby we can obtain Taylor polynomials in time derivatives of different functions of the external force, e.g. for the equation (26) there are three possibilities: $k_1, k_3 > 0$, $k_1 = 0$ , or $k_3 = 0$. The resulting LT-formulas provide information about the significance of individual components of PO for its LT-dynamics !

### 4.3 General relativistic properties of the LT-dynamics

According to the hypothetical relativistic LT-formula (36), the LT-dynamics of a PO is specified up to the order of $\lambda^3$ inclusive by the four LT-constants $k_1$, $k_2$, $k_{31}$ and $k_{32}$. According to Einstein, the first LT-constant is $k_1 = 1/mc$ . In addition, within Dirac's classical theory of radiating electrons, the second LT-constant is $k_2 = q^2/6\pi\epsilon_0 mc^3$. Therefore, in general we expect that the LT-constant $k_2$ is determined by the intensity of the loss of the PO four-momentum in response to the time derivative of the external force, it is not negative as a real PO may provide only a finite amount of the four-momentum. Therefore, the first two terms of the relativistic LT-formula (36) provide for a classical electrified PO the definitions of its mass and charge by their role in the relativistic LT-dynamics.

The physical interpretation of two third order LT-constants $\mathrm{k}_{31}$ and $\mathrm{k}_{32}$ is open. Under the Bhabha condition (40, they are related as follows:

$$\mathrm{k}_{31} = (1 - \tfrac{2}{3}c_1){k_1}^{-1}{k_2}^2 \qquad \text{and} \qquad \mathrm{k}_{32} = (1 - c_1){k_1}^{-1}{k_2}^2 \,, \tag{44}$$

where $c_1$ is a real constant.

Using the LT-formula (36) with increasing $\lambda \geq 0$, we can simulate how at a given precision of our observations the number of the observable LT-constants increases with the magnitude and rate of change of the external four-force $\lambda\Phi(\lambda \mathrm{t})$ : at very small and slowly changing external force, just the PO-mass can be determined. Due to the multitude of actual POs, we see no physical reason to believe that there is only a limited number of the independent relativistic LT-constants.

### 4.4 Application of the LT-formula

**1)** The N-formulas (17) and (18) point out exceptional cases with POs whose LT-formula equals their N-formula.

**2)** As we can hardly ever obtain the exact solutions in closed form for the PO equation of motion, the corresponding LT-formulas provide a welcome source of information about LT-dynamics. Whenever each following term of the LT-formula is essentially smaller than the preceding one, we may expect that this formula will provide appropriate information about the LT-dynamics. The LT-constants provide the significance of the individual kinetic constants of the PO equation of motion for LT-dynamics.

**3)** A hypothetical LT-equation implies hypothetical LT-formula of the same order of $\lambda$: We may always use it to calculate approximations to the long-term PO-trajectories because LT-formula cannot exhibit self-acceleration, on the contrary to PO-equation

**4)** Certain PO moving through a medium might present a too complicated system to create its exact equation of motion. But we may still be able to somewhat describe its LT-dynamics by a hypothetical LT-equation which balances the external force with the sum of the inertial force and a S-model of the interaction force between PO and the surroundings, cf. the hypothetical LT-equation (30) and the Lorentz-Abraham-Dirac equation (41).

**5)** To make a hypothetical LT-formula we can use as a generic ansatz some multivariate polynomials in time derivatives of an appropriate function of the external force such as given by the equations (19), (29), (32) or (36), and extract the values of their parameters by multiple linear regressions from data about acceleration the of long-term PO-trajectories, see [6]. Thus without knowing an adequate N-formula, we can make appropriate LT-formulas for predicting the LT-dynamics of a given PO.

**6)** There are many real systems consisting of POs, each of which is treated as the point mass with acceleration specified by Newton's second law, e.g. in astronomy, and in classical mechanics. Using appropriate LT-formulas instead of Newton's second law, we could take into account not only the PO-masses but also some of the additional kinetic properties of these POs. That way we might get better dynamic models of such systems.

**7)** Basing models of continuous mechanical medium on the laws about the interaction of the point masses and Newton's second law (see [7, Sect. 4.4]), and using an appropriate LT-formula instead of Newton's second law, one might get better models.

**8)** Remaining subjects: i) Memory: PO-acceleration after cessation of the external force, ii) Passive damping: energy dissipation by parts of a composite PO.
iii) Composite dynamics: the LT-dynamics of a PO in terms of the LT-formulas of its parts.

## B. Summary

We presented a particular type of simplified models. Using few terms of the Taylor series, these models provide some information about difficult and complex problems. In this paper, they describe the long-term dynamics of a special type of bodies at small and slowly changing external forces. We introduced:

**1.** A pointlike physical object PO, a classical extended object whose motion interacts with the sourrounding medium. PO response to an external force is aptly specified solely by the trajectory of a single point, whose velocity eventually stops changing after the cessation of the external force.

**2.** A particular type of PO-dynamics: if the PO-acceleration is cyclic then LT-dynamics equals the basic dynamics, there is no start-up.

**3.** **The key mathematical** element is the N-formula. It generalizes Newton's second law formula for acceleration by explicitly specifying the PO-acceleration by the external force.

**4.** The simplified models of the LT-dynamics under a small and slowly changing external force $\lambda F(\lambda t)$, i.e. the LT-formulas that approximate the long-term dependence of PO-acceleration at a given time instant by the Taylor polynomial in time derivatives of the external force at the same time instant.

Given an ordinary PO differential equation of motion, we can calculate *iteratively* the corresponding LT-formula of any order of $\lambda$! Nevertheless, the LT-formula does not need to imply the original PO equation of motion.

Each LT-formula implies iteratively a novel differential equation of the same order of $\lambda$ about the LT-dynamics (LT-equation), and vice versa. Such an LT-equation provides certain information about the LT-dynamics, and may exhibit self-acceleration. Different POs may have identical LT-formulas of the same order of $\lambda$.

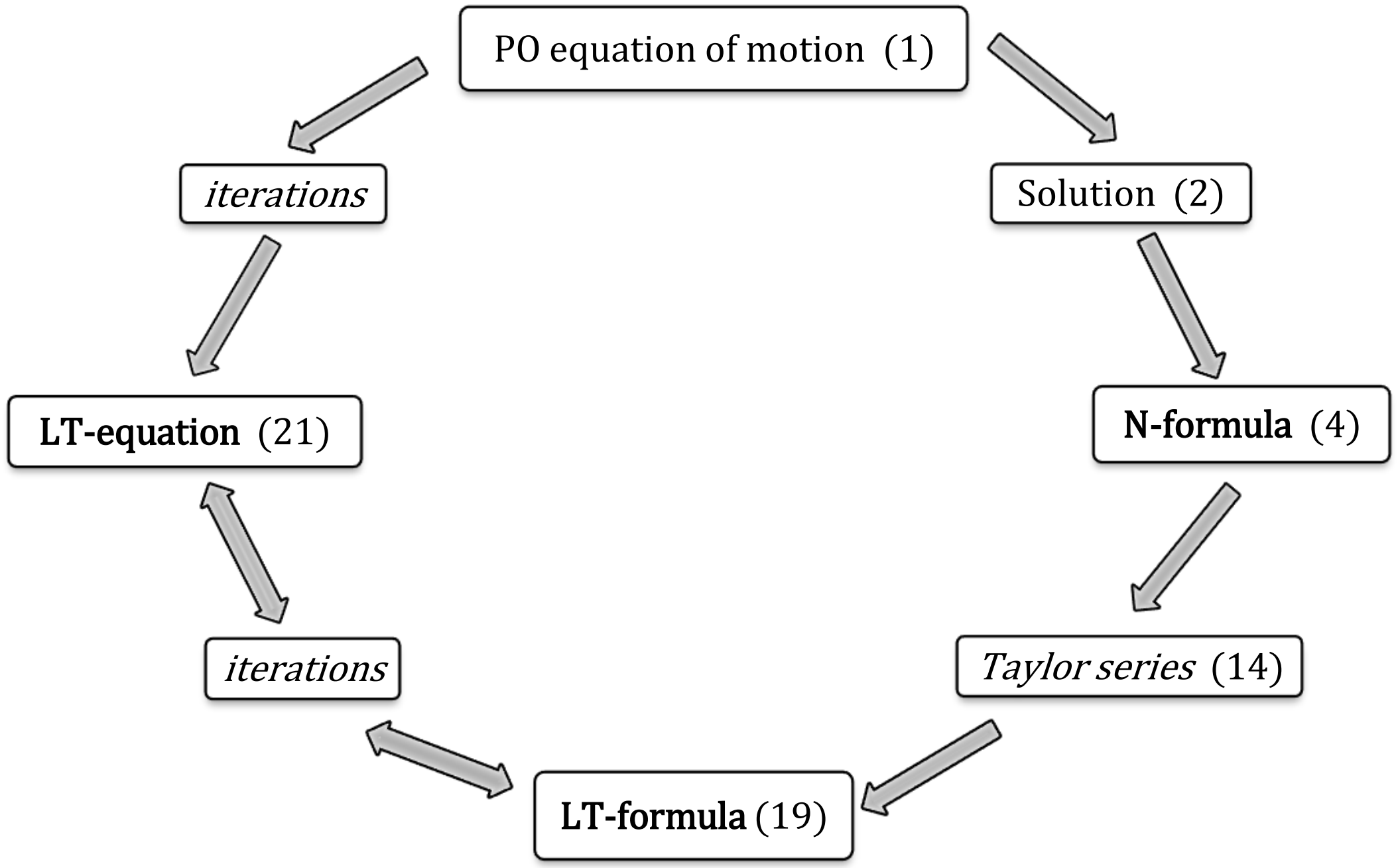

**Figure 1: The simplified modeling of the long-term PO dynamics**

To illustrate the mathematical framework and usage of serial models of the long-term dynamics of the pointlike objects, we considered the driven damped harmonic oscillator in Section 2. As Figure 1 the block cycle diagram above shows, we illuminated the relations between various elements and techniques for modeling the PO long-term dynamics, where t∈ $(t_m, +\infty)$ and $t_m \nearrow +\infty$, i.e. $t_m$ is large. Thus, the PO long-term dynamics is actually a generalization of the asymptotic dynamics, where t ↗ +∞.

## C. Main points

**(i) Simplification through Taylor series expansion.** We consider simplification of mathematical models by Taylor series expansion of their elements. According to Planck [16] " the simpler the presentation of a particular law of Nature, the more general it is though at the same time, which formula to take as the simpler, is a problem which cannot always be confidently and finally decided."

**(ii) Modifications of an initial ansatz through its parameter.** The tacit basis of our approach to model making is the fact that *when the given ansatz model is an invertible function of a particular parameter; we can modify it through this parameter into an arbitrary ansatz.*

**Example**: Given the three parameter ansatz

$$F(x;\alpha,b,c) \equiv (a+bx)/(1+cx),\ x,a,b,c \in (-\infty.+\infty)\,, \qquad (45)$$

then for any function $f(x)$ we get

$$F(x;\alpha,b,c_f) = f(x) \qquad (46)$$

iff

$$c_f \equiv (a+bx-f)/(xf)\,. \qquad (47)$$

**(iii)** To model small and slowly changing force we use $\lambda F(\lambda t)$ with small variable $\boldsymbol{\lambda} > 0$. The results are useful with $\lambda = 1$ if function $F(t)$ in itself is small and slowly changing**.**
**(iv)** We introduced **pointlike objects** and **serial models** as the modeling elements. To illustrate them we provided various examples of heuristic approach to modeling the pointlike objects dynamics. This might prevent future mix-up over the proper usage of the long-term equations and various differential equations for modeling dynamics of the pointlike objects. We put forward this novel approach as well we knew, yet

"Proof of the pudding is in eating".

**Acknowledgement**

We express our thanks to S. Ribarič and to the JAMP reviewer for many helpful comments.

❖ *We appreciate any comments, references, suggestions, and viewpoints.*

**5. References**

[1] G. E. Hrabovsky, Introduction to Theoretical Physics,(2000) http://www.madscitech.org/notes/series1/day1. pdf
[2] R. P. Feynman, The Development of the Space-Time View of Quantum Electrodynamics, Nobel Lecture, 1965. https://www.nobelprize.org/nobel_prizes/physics/laureates/1965/html.
[3] M. Ribarič and L. Šušteršič, A new window into the properties of the Universe: Modification of the QFTs so as to make their diagrams convergent, **arXiv:1503.06325v3** [hep-th**]** (2016).
[4] M. Ribarič and L. Šušteršič, Empirical relationship as a stepping-stone to theory, arXiv:0810.0905 [physics.gen-ph] (2021).
[5] P. A .M. Dirac, Science Quotes by Paul A. M. Dirac, (1984) http://todayinsci.com/D/Dirac_Paul/DiracPaul-Quotations.htm
[6] G. A. Schott, On the motion of the Lorentz electron *Phil. Mag. S.6*. **29** 49-62 (1915).
[7] M. Ribarič and L. Šušteršič, *Conservation Laws and Open Questions of Classical Electrodynamics* (Singapore: World Scientific 1990) PDF copy Marijan Ribarič (researchgate.net)/Research
[8] P. A. M. Dirac, Classical theory of radiating electrons *Proc. Roy. Soc. (London)* **A167** 148-169 (1938).
[9] H. J. Bhabha, Classical theory of electrons *Proc. Indian Acad. Sci.* **A10** 324-332 (1939).
[10] M. Ribarič and L.Šušteršič, A differential relation for slowly accelerated point like charged particles *Phys. Lett.* **A139** 5-8 (1989) .PDF copy Marijan Ribarič (researchgate.net)/Research
[11] M. Ribarič and L. Šušteršič, Improvement on the Lorentz-Abraham-Dirac equation arXiv: 1011.1805v2 [physics.gen-ph] (2010).
[12] F. Rohrlich, Dynamics of a charged particle *Phys. Rev*. **E77** (2008) 046609 arXiv:0804.4614(2008).
[13] N. M. Naumova and I. V. Sokolov, Comment on "Dynamics of a Charged Particle" by F. Rohrlich [Phys. Rev. E 77, 046609 (2008), ] arXiv:0904.2377 (2009).

[14] F. Rohrlich, *Classical Charged Particles* (Singapore: World Scientific 2007).
[15] R. F. O'Connell, Radiation reaction: general approach and applications, especially to electrodynamics *Contemporary Physics* **53**(4) 301-313 (2012).
[16] Max Planck, The Genesis and Present State of Development of the Quantum Theory, Nobel Lecture, 1920 https://www.nobelprize.org/prizes/physics/1918/planck/lecture.